\documentstyle[twocolumn,12pt]{article}
\begin{document}
\baselineskip=11pt
\parskip=0pt
\onecolumn
\begin{center} 
{\Large\bf 
Quantum Theory for Mesoscopic Electronic\\ 
Circuits and Its Applications 
\footnote{Talk on the 3rd IMACS/IEEE international multi-conference:
{\it CSCC'99}, Athens}
} 
\bigskip

You-Quan Li
\footnote{email: yqli@Physik.Uni-Augsburg.DE}\\
Zhejiang Institute of Modern Physics\\
Zhejiang University\\ 
Yugu Road 20, Hangzhou 310027\\
CHINA\\

\bigskip

Bin Chen\\
Institute of Theoretical Physics\\
Chinese Academy  of Science\\ 
P.O. Box 2735, Beijing 100080\\ 
CHINA\\
\bigskip
(February 26, 1999)

\bigskip\bigskip\bigskip
\end{center}

{\it Abstract:}
This talk contains an overview of the quantum theory for mesoscopic
electric circuits and some of its applications.
In the theory, the importance of the discreteness of electronic charge
in mesoscopic electric circuit is addressed. The mesoscopic LC-design
is quantized in accord with the charge discreteness. The uncertainty
relation for electric charge and current is obtained.
Because the stationary Schr\"odinger equation is turned
to be Mathieu equation in $p$-representation, the wave function and energy
spectrum is formally solved.

As further applications, the persistent current is obtained by considering
the mesoscopic ring as a pure L-design. 
A formula for
persistent current arising from magnetic flux is obtained from a new point
of view.
The Coulomb blockade phenomenon
occurs when we applying the theory to the pure C-design. Concerning the time
evolution of the state for mesoscopic electric circuit, we are able to study
it in terms of the method of characteristics. 
%In a L-design, some kind of Bloch oscillation occur. 
In order to study the dissipative effect in
the circuit, we use density-matrix formulation. 
In this formulation, several type of ``off diagonal'' dissipations 
%are discussed extensively.
are expected to be discussed.

\bigskip
{\it Key-Words:} Mesoscopic circuit, discreteness of charge, quantization,
Coulomb  Blockade, persistent current
%\vfill
\twocolumn
\section{Introduction}

Along with the dramatic achievement in nanotechnology,
such as molecular beam epitaxy, atomic scale fabrication or advanced
lithography, mesoscopic physics and nanoelectronics are
undergoing a rapid development\cite{IBM,Buot}.
It has been a strong and definite trend in the miniaturization
of integrated circuits and components towards atomic-scale
dimensions\cite{Garcia} for the electronic device community.
When transport dimension reaches a characteristic dimension,
namely, the charge carrier inelastic coherence length, one must address
not only  quantum mechanical property but also
the discreteness of electron charge. Thus a correct quantum
theory is indispensable for the device physics in integrated circuits of 
nanoelectronics. 
The quantization of the circuit was carried out
\cite{Louisell} just likes that of a harmonic oscillator.
This only results energy quantization.
In fact, a different kind of fluctuation in
mesoscopic system
%, which inherently has nothing to do with energy
%quantization and interference
%of wave functions,
is due to the quantization of electronic charge. 
We studied the quantization of electric circuit of LC-design
under the consideration of the discreteness of electric charge\cite{LiCh1}.

In this talk, we demonstrate the quantum theory for mesoscopic 
electric circuits and it applications.
After recalling the Kirchoff's law in classical LC-design in next 
section, we quantize the LC-design by considering the fact of 
charge discreteness in section \ref{sec:quantization}, where
a finite-difference Schr\"{o}dinger equation for the mesoscopic electric
circuit is proposed.
In section \ref{sec:uncertainty},
the uncertainty relation for charge and current
is discussed. 
In section \ref{sec:LC},
the finite-difference Schr\"{o}dinger equation for a mesoscopic
circuit of LC-design is turned to  Mathieu equation in
`p-representation' and solved exactly. 
In section \ref{sec:L}, 
A gauge field
is introduced and a formula for persistent current which is a periodic
function of the magnetic flux is obtained. 
In section \ref{sec:C}, The pure C-design is studied and the Coulomb 
blockade is formulated. 
In section \ref{sec:time}, we study the time evolution of the system, 
particularly
the L-design in the presence of time-dependent source. 
In section \ref{sec:dissipative},
we suggest an approach to include the dissipative effects in the circuit.
In section \ref{sec:discussion}, we summarize the main results and give some 
discussions.

\section{The classical LC-design}\label{sec:classical}

A classical  non-dissipative electric 
circuit of LC-design fulfills the 
Kirchoff's law, {\it i.e.}

\[
\dot{q} = \frac{\partial H }{\partial p }, \,\,
\dot{p} = - \frac{\partial H }{ \partial q }
\]
with $H(t) = {\displaystyle
\frac{ 1 }{ 2L } p^2 + \frac{ 1 }{2C}q^2 + \varepsilon(t) q
} $ as the Hamiltonian.
The variable $q$ stands for the electric charge instead of
the conventional `coordinate', while its conjugation 
$
{\displaystyle
p(t) = L dq /d t
} $
represents (apart from a factor L ) the  electric current instead of
the conventional `momentum'.

\section{Quantization} 
\label{sec:quantization}

In order to take account of the discreteness of 
electronic charge. 
we must impose that
the eigenvalues of the self-adjoint operator
$\hat{q}$ take discrete
values\cite{LiCh1}, i.e.
\begin{equation}
\hat{q} | q > = n q_e | q >
\label{eq:c}\end{equation}
where $n \in {\sf \, Z \!\!\! Z \, } $ (set of integers) and
$q_e  = 1.602 \times 10^{-19}$ coulomb,
the elementary electric charge.  Obviously, any eigenstate of
$\hat{q}$ can be specified by an integer. This allows us to
introduce a minimum `shift operator'
$
\hat{Q} := \exp [iq_e \hat{p} /\hbar],
$
which  satisfies 
\begin{eqnarray}
[ \hat{q}, \hat{Q} ] = - q_e \hat{Q}, \;\;
[ \hat{q}, \hat{Q}^{+} ] = q_e \hat{Q}^{+}, \nonumber \\[2mm]
\hat{Q}^{+} \hat{Q} = \hat{Q} \hat{Q}^{+} = 1.\hspace{10mm}
\label{eq:d}\end{eqnarray}
In ref.\cite{LiCh1} we 
obtained the following
finite-difference Schr\"{o}dinger equation,
\begin{equation}
\left[
-\frac{\hbar^2}{2q_e L}(\nabla_{q_e }-\bar{\nabla}_{q_e })
+V(\hat{q})
\right]|\psi>=E|\psi>.
\label{eq:h}\end{equation}
This is the stationary equation, the time evolution will be studied in
section \ref{sec:time}.

\section{Uncertainty relation}
\label{sec:uncertainty}

We have shown \cite{LiCh1} that 
the uncertainty relation for electric charge and electric current,
becomes
\begin{equation}
\Delta\hat{q}\cdot\Delta\hat{P} \geq \frac{\hbar}{ 2 }
( 1 + \frac{q^2_e }{\hbar^2} < \hat{H}_0 > ).
\label{eq:de}
\end{equation}
Obviously, this uncertainty relation recovers the usual Heisenberg
uncertainty relation if $ q_e $ goes to zero, i.e. the case that the
discreteness of electric charge vanishes. Moreover, the uncertainty relation
(\ref{eq:de}) has shown us further knowledge than 
the traditional
Heisenberg uncertainty relation.

\section{The quantum LC-design}\label{sec:LC}

We only consider the adiabatic approximation here so that
$\varepsilon (t)$ is considered as a constant $\varepsilon$,
{\it i.e.}, $V(\hat{q})=C^{-1}\hat{q}^2/2 + \varepsilon\hat{q}$.
Let us consider a representation in which the operator $\hat{p}$ is
diagonal and called it as p-representation. 
The transformation of wave functions between charge representation
and p-representation is given by
\begin{equation}
< n | \psi > = (\frac{q_e}{2\pi\hbar} )
\int^{\hbar(\frac{\pi}{q_e}) }_{-\hbar(\frac{\pi}{q_e} ) }
dp <p |\psi >
e^{-inq_e p /\hbar}
\label{eq:pton}\end{equation}

In the `p-representation', the finite-difference  Schr\"{o}dinger
equation becomes a differential equation for
$\tilde{\psi}(p):=<p|\psi>$,
\begin{equation}
\left[
- \frac{\hbar^2 }{2C} \frac{\partial^2 }{\partial p^2 }
- \frac{\hbar^2}{q_{e}^2 L }( \cos(\frac{q_e}{\hbar} p ) - 1 )
\right] \tilde{\psi } (p) = E\tilde{\psi } (p).
\label{eq:l}\end{equation}
which is the well known Mathieu equation \cite{Wang,Grad}. This equation
was ever appeared in \cite{Jurk} on the discussion of Pad\'{e} approximates.

In terms of the conventional notations \cite{Wang,Grad}, the wave
functions in p-representation can be solved as follows
$$
\tilde{\psi}^{+}_{l}(p) =
{\rm ce }_l (\frac{\pi}{2}- \frac{q_e }{ 2\hbar}p , \,\xi )
$$
or
\begin{equation}
\tilde{\psi}^{-}_{l+1}(p) =
{\rm se}_{l+1}(\frac{\pi}{2}-\frac{q_e}{2\hbar}p,\,\xi)
\label{eq:m}
\end{equation}
where the superscripts `+' and `-' specify  the even and odd parity
solutions respectively;
$l = 0, 1, 2, \cdots $;
$\xi = (2\hbar/q^{2}_e )^2
C/L$; ce$(z, \xi)$ and se$(z, \xi)$ are periodic Mathieu
functions.  In this case, there exist infinitely many eigenvalues
$\{ a_l \}$ and $\{ b_{l+1} \} $ which are not identically equal to zero.
Then the energy spectrum is expressed in terms of the
eigenvalues $a_l $, $b_{l}$ of Mathieu equation

\begin{eqnarray}
E^{+}_l =  \frac{q_{e}^2 }{8C}a_l(\xi)
+ \frac{\hbar^2 }{ q_{e}^2 L},
\nonumber \\
E^{-}_{l+1} =  \frac{q_{e}^2}{8C}b_{l+1}(\xi)
+ \frac{\hbar^2 }{ q_{e}^2 L}.
\label{eq:n}\end{eqnarray}

\section{Quantum L-design and persistent currents }
\label{sec:L}

Introducing an operator
$
\hat{G} :=\exp(-i\beta\hat{q}/\hbar)
$, 
we can find that
$\hat{G} | p > = | p - \beta >$
and
$\hat{G}^+ | p > = | p + \beta >.$
For a unitary transformation to the eigenstates of
Schr\"{o}dinger operator given by
\[
| \psi > \rightarrow  | \psi' > = \hat{G} | \psi >,
\]
we find that the Schr\"{o}dinger equation (\ref{eq:h}) is not covariant.
This requests that we introduce a gauge field and define a reasonable
covariant discrete derivative. By making the definitions:
\begin{eqnarray}
D_{q_e} := e^{-i(q_e/\hbar)\phi }
     \frac{ \hat{Q} - e^{ i(q_e/\hbar)\phi } }
          { q_e } , \nonumber \\
\bar{D}_{q_e} := e^{i(q_e/\hbar)\phi }
\frac{ e^{-i(q_e/\hbar)\phi} - \hat{Q^+}}
     { q_e } ,
\label{eq:eg}
\end{eqnarray}
we can verify that they are covariant under a gauge transformation. The
gauge transformations are expressed as
\begin{eqnarray}
\hat{G} D_{q_e} \hat{G}^{-1} = D'_{q_e}, \nonumber \\
\hat{G} \bar{D}_{q_e} \hat{G}^{-1} = \bar{D}'_{q_e},
\label{eq:eh}
\end{eqnarray}
as long as the gauge field $\phi$ transforms in the following way
\[
\phi\rightarrow\phi'=\phi-\beta.
\]
Either the transformation law or the dimension of the field $\phi$
indicates that $\phi$ plays the role of the magnetic flux threading
the circuit.
Then the Schr\"{o}dinger equation for
a pure L-design in the presence of magnetic flux is given by,
\begin{equation}
-\frac{\hbar^2 }{2q_e L } (D_{q_e}\, - \bar{D}_{q_e} )
| \psi > = E | \psi >.
\label{eq:gh}
\end{equation}
The energy spectrum is easily solved
\begin{equation}
E(p,\phi) = \frac{2\hbar}{q^2_e} \sin^2
\left[\frac{q_e}{2\hbar}(p-\phi)
\right],
\end{equation}
which has oscillatory property with respect to
$\phi$ or $p$. Differing from the usual classical pure L-design, the
energy of a mesoscopic quantum pure L-design can not be large than
$2\hbar / q_e^2 $.
Clearly, the lowest energy states are such states that
$p = \phi + n h/q_e$.
Thus the eigenvalues of the electric current 
of ground state are calculated
\begin{equation}
I(\phi) = \frac{\hbar}{q_e L} \sin( \frac{q_e}{\hbar} \phi ).
\label{eq:pc}
\end{equation}
Obviously, the electric current on a mesoscopic  circuit of pure L-design
is not null in the presence of a magnetic flux except
$\phi = n(h/q_e)$.
Clearly, this is a pure quantum characteristic.
(\ref{eq:pc}) exhibits that the persistent current in a mesoscopic L-design
is an observable quantity periodically depending on the flux $\phi$.
Because a mesoscopic metal ring is a natural pure L-design, the formula
(\ref{eq:pc}) is valid for persistent current on a single mesoscopic
ring\cite{Chand}. One can easily calculate the inductance of mesoscopic
metal ring and obtain
the formula for persistent currents
\begin{equation}
I(\phi) = \frac{\hbar}{8\pi r 
   (\frac{1}{2}\ln(8r/a) - 1)q_e}
     \sin(\frac{q_e}{\hbar}\phi),
\end{equation}
where $r$ is the radius of the ring and $a$ is the radius of the metal wire.
Differing from the conventional formulation of the persistent
current on the basis of quantum dynamics for electrons, our formulation
presented a  method from a new point of view. Formally, the $I(\phi)$
we obtained is a sine function with periodicity of
$\phi_0 =h/q_e$, 
But either the model that the electrons move freely
in an ideal ring\cite{Ch1}, or the model that the electrons have hard-core
interactions between them\cite{LiMa2} can only give the sawtooth-type
periodicity. 

\section{Quantum C-design and Coulomb blockade}\label{sec:C}

We observe the Schr\"{o}dinger equation for a LC-design.
The mesoscopic capacity may be relatively very small 
(about $10^{-8}F$) but the inductance of a macroscopic circuit 
connecting to an adiabatic voltage source is relatively large
because the inductance of a circuit is proportional to the
area which the circuit span. Thus we may neglect the term 
reversely proportional  to $L$ in the Schr\"{o}dinger equation
and study the equation for a pure C-design:
\begin{equation}
\left(\frac{1}{2C}\hat{q}^2 -\varepsilon\hat{q}
  \right)|\psi>=E|\psi>.
\end{equation}
Apparently, the Hamiltonian operator commutes with 
the charge operator,
so they have simultaneous eigenstates. 
The energy for the eigenstate $|n>$ is 
\begin{equation}
E=\frac{1}{2C}(n q_e - C\varepsilon)^2 
  -\frac{C}{2}\varepsilon^2,
\end{equation}
which involves both the charge quantum number and the voltage source.
After some analysis, we can find the relations between charge $q$
and the voltage $\varepsilon$ for the ground state,
\begin{eqnarray}
q=\sum_{m=0}^{\infty}
   \left\{ \theta[\varepsilon -(m+\frac{1}{2})\frac{q_e}{C}]\right.
              \nonumber\\  
            \left. -\theta[-\varepsilon -(m+\frac{1}{2})\frac{q_e}{C}]
              \right\}q_e.
\label{eq:chargestep}
\end{eqnarray}
where $\theta(x)$ denotes the step function.
The corresponding eigenstates are
\begin{eqnarray}
|\psi(\varepsilon)>_{ground}=
   \sum_{m=-\infty}^{\infty}
     \left\{\theta[\varepsilon -(m-\frac{1}{2})\frac{q_e}{C}]
            \right. \nonumber \\
            -\left.\theta[-\varepsilon -(m+\frac{1}{2})\frac{q_e}{C}]
             \right\}|m>
\end{eqnarray}
The dependence of the currents on the time is obtained by taking 
derivative of (\ref{eq:chargestep}),
\begin{eqnarray}
\frac{dq}{dt}=\sum_{m=0}^{\infty}q_e
   \left\{\delta[\varepsilon -(m+\frac{1}{2})\frac{q_e}{C}]
             \right.    \nonumber\\    
            \left. +\delta[\varepsilon + (m+\frac{1}{2})\frac{q_e}{C}]
              \right\}\frac{d\varepsilon}{dt}.
\end{eqnarray}
Clearly, the currents are of the form of sharp pulses which occurs
periodically according to the changes of voltage. The voltage
difference between two pulses are $q_e/C$. This is the called Coulomb
blockade phenomena caused by the charge discreteness. 

\section{Time evolution 
with time dependent source}\label{sec:time}

We consider a L-design in the presence of time dependent source. The time
dependent Schr\"odinger equation for the circuit in charge representation is 
given by
\begin{eqnarray}
i\hbar\frac{d}{dt}|\psi(t)>=\hspace{38mm}\nonumber\\
\left[
   -\frac{\hbar^2}{2q_eL}(\nabla_{q_e}-\bar\nabla_{q_e})
     +\varepsilon(t)\hat{q} \right]|\psi(t)>.
\label{eq:timeSch}
\end{eqnarray}
Expanding the state vector in terms of the orthonormal set of charge eigenstates
\begin{equation}
|\psi(t)>=\sum_{n=-\infty}^{\infty}u_n(t)|n>,
\end{equation}
we obtain the evolution equation for the amplitudes $u_n(t)$, namely
\begin{eqnarray}
i\frac{d}{dt}u_n(t)=[n\frac{q_e}{\hbar}\epsilon(t) +\frac{\hbar}{q_eL}]
   u_n(t)  \nonumber\\  
-\frac{\hbar}{2q_eL}(u_{n+1}(t)+u_{n-1}).
\label{eq:timeevo}
\end{eqnarray}
Obviously, $du_n/dt=-i\omega u_n$ corresponds to the stationary case which
was discussed in section \ref{sec:L}. 

Multiplying eq.(\ref{eq:pton}) with $\exp(inq_ep'/\hbar)$ and 
summing over the index $n$, we obtain
\begin{equation}
<p|\psi>=\sum_{n=-\infty}^\infty <n|\psi>e^{inq_e p/\hbar}.
\end{equation}
Then eq. (\ref{eq:timeSch}) can be written in a convenient form 
in $p$-representation, namely,
\begin{equation}
\left\{-\frac{\hbar}{q_eL}[\cos(\frac{q_e}{\hbar}p)-1]
 -i\frac{\hbar}{q_e}\varepsilon(t)\frac{\partial}{\partial p}\right\}
   \tilde\psi
   =i\frac{d}{dt}\tilde\psi,
\label{eq:partialeq}
\end{equation}
where
\begin{equation}
\tilde\psi(p)=<p|\psi> 
=\sum_{n=-\infty}^\infty u_n(t)e^{inq_ep/\hbar}.
\end{equation}
The partial differential equation (\ref{eq:partialeq}) can be reduced to a first
order ordinary differential equation by means of the method of characteristics
\cite{character}. Although this reduction involves complicated mathematical
formulations, the solution of the above equation can be given in terms of 
Bessel functions\cite{Dunlap}.
Using Graf's addition theorem for Bessel functions
\cite{Grad}, one is able to obtain 
\begin{equation}
|u_n(t)|^2 =J^2_n(-\frac{\hbar}{q_e L}\sqrt{a^2(t)+b^2(t)}\,),
\end{equation}
where
\begin{eqnarray*}
a(t)&=&\int^{t}_0 \cos[f(t')]dt', \\
b(t)&=&-\int^{t}_0 \sin[f(t')]dt' \\
f(t)&=&\int^{t}_0 \varepsilon(t')dt'.
\end{eqnarray*}
The mean-square charge is obtained with the help of 
the identity $\sum_n J_n^2(z) = z^2/2$,
\begin{equation}
<\hat{q}^2>=\frac{\hbar^2}{2L}[a^2(t)+b^2(t)].
\end{equation}
Some interesting physics phenomena, such as dynamic
localization etc are expected to be studied furthermore. 
 
\section{About the dissipatives}
\label{sec:dissipative}

In order to take account of the effects of dissipatives in the circuit, we
introduce the density matrix $\rho_{m,n}(t)=u^*_m(t) u_n(t)$.
The evolution equation for the density matrix is obtained from 
the equation (\ref{eq:timeevo}) for the amplitudes $u_n(t)$,
\begin{eqnarray}
i\frac{\partial}{\partial t}\rho_{m,n} =
  -\varepsilon(t)\frac{q_e}{\hbar}(m-n)\rho_{m,n}\hspace{12mm}\nonumber\\
   +\frac{\hbar}{2q^2_eL}(\rho_{m+1,n}+\rho_{m-1,n}
        -\rho_{m,n+1}-\rho_{m,n-1}).
\end{eqnarray}
Putting a term describing some kind of dissipatives in the circuit, we should
write the evolution equation for the density matrix as
\begin{eqnarray}
i\frac{\partial}{\partial t}\rho_{m,n}=
  -\varepsilon(t)\frac{q_e}{\hbar}(m-n)\rho_{m,n}
                    \nonumber\\
   +\frac{\hbar}{2q^2_eL}(\rho_{m+1,n}+\rho_{m-1,n}
        -\rho_{m,n+1}-\rho_{m,n-1})    \nonumber\\
   +i\sum_{m',n'}\gamma^{m' n'}_{m n}\rho_{m' n'},\hspace{1cm}
\label{eq:SLiouville}
\end{eqnarray}
which is the stochastic Liouville equation. Eq.(\ref{eq:SLiouville})
is a good start point to study the dissipative effects in the mesoscopic
L-design. The one-parameter off-diagonal dissipative is particularly 
interesting, {\it i.e.}, 
$
\gamma_{m' n'}^{m n}=\gamma (1-\delta_{m,n})\delta_{m,m'}\delta_{n,n'}
$
Further discussions are in progress.

\section{Conclusions}
\label{sec:discussion}

In the above, we studied the quantization of  mesoscopic electric
circuit. Differing from the literature in which it is simply treated
as the quantization of a harmonic oscillator, we addressed the importance
of the discreteness of electric charge. Taking the discreteness into
account, we proposed a quantum theory for mesoscopic electric circuit
and give a finite-difference Schr\"{o}dinger equation for mesoscopic
electric circuit.
We used the charge representation and
p-representation. Due to the discreteness of electric
charge, $\hat{p}$ is no longer a current operator.

As the Schr\"{o}dinger equation for LC-design in
p-representation becomes the well known Mathieu
equation, it is solvable. We obtain the wave functions
in terms of Mathieu functions and the energy
spectrum in terms of the eigenvalues of Mathieu equation.
The discussion on uncertainty relation for the charge and current shed some
new light on the knowledge of transitional Heisenberg uncertainty relation.
As further applications  of our theory,
Introducing a gauge field and gauge transformation,
we obtained a formula for the persistent current on
the mesoscopic pure L-design in the presence of the magnetic flux.
As the mesoscopic metal ring is a natural pure L-design,
the formula is certainly valid for the persistent current on
mesoscopic rings.
In our formula, the mass of electrons, the carriers for electric current,
is not involved. It is worthwhile to check that property  by experiments.
The  theory is applied to explain the Coulomb blockade. 
By considering a quantum C-design, the Coulomb blockade phenomenon was 
formulated as an immediate 
consequence. The time evolution of quantum L-design is solved formally, 
and the dissipative is introduced with the help of density matrix formulation.

%\section*{Acknowledgment}
The work is supported by NSFC No. 19675030 and NSFZ No.198024.

\end{document}